\documentclass[aps,prb,onecolumn,groupedaddress]{revtex4}
\usepackage{graphicx}

\begin{document}
\title{ Angular Dependent Torque Measurements on CeCoIn$_5$ Single Crystals}
\author{H. Xiao, T. Hu, C. C. Almasan}
\affiliation{Department of Physics, Kent State University, Kent, Ohio, 44242, USA}
\author{T. A. Sayles, M. B. Maple}
\affiliation{Department of Physics, University of California at San Diego, La Jalla, California, 92903, USA}

\date{\today}
\begin{abstract}
Angular dependent torque measurements were performed on single crystals of CeCoIn$_5$, a heavy fermion superconductor 
($T_{c0}$ = 2.3 K), in 
the temperature, $T$, range 1.9  K $\leq T \leq $ 20  K and magnetic fields $H$ up to 14 T. A large paramagnetic effect 
is found in the normal state. Torque measurements in the mixed state were
also performed. The torque curves show sharp hysteresis peaks at $\theta = 90^{o}$ ($\theta$ is the angle between
$H$ and the
$c-$axis of the crystal), a result of 
intrinsic pinning of vortices. The anisotropy 
$\gamma \equiv \sqrt{m_{c}/m_{a}}$ in the mixed state was determined from the reversible part of the vortex contribution 
to the torque signal 
using Kogan's model
[Phys. Rev. B \textbf{38}, 7049 (1988)]. The anisotropy $\gamma$ decreases with increasing magnetic field and temperature. 
The fact that $\gamma$ is not 
a constant points towards a multiband scenario in this heavy fermion material.
\end{abstract}


\maketitle

\subsection{Introduction}
The heavy fermion compound CeCoIn$_5$ forms in the HoCoGa$_5$ tetragonal crystal structure with alternating layers of 
CeIn$_3$ and CoIn$_2$. It is superconducting at 2.3 K, the highest superconducting transition
temperature $T_{c0}$ yet reported for a heavy fermion superconductor. \cite{Petrovic} Considerable progress has been made in 
determining
the physical properties of this material. The superconductivity in
 this material is unconventional. The presence of a strong magnetic interaction between the  4f moments and itinerant electrons 
allows the 
possibility of nonphonon mediated coupling between quasiparticles. \cite{Movshovich, Mathur} Angular dependent thermal conductivity measurements show $d_{x^{2}-y^{2}}$ symmetry which implies that 
the anisotropic antiferromagnetic
fluctuations play an important role in superconductivity.\cite{Izawa} Bel et al. reported the giant Nernst effect in the 
normal state of CeCoIn$_5$, which is
comparable to high $T_c$ superconductors in the superconducting state. \cite{Bel} Non-Fermi liquid behavior was observed 
in many aspects. \cite{Sidorov, Kim,
Petrovic} The unconventional superconductivity and the similarity to high $T_{c}$ superconductors attract great interest 
to study this system.

Measurements of de Haas-van Alphen oscillations in both the normal and mixed states have revealed the quasi two-dimensional 
nature of the Fermi surface and the presence of a small number of electrons exhibiting 3D behavior.\cite {Settai, Hall} Through point-contact spectroscopy
measurement,  Rourke et al.\cite {Rourke} found that in CeCoIn$_5$ there are two coexisting order parameter components with
amplitudes $\Delta_1$ = 0.95 $\pm$ 0.15 meV and $\Delta_2$ = 2.4 $\pm$ 0.3 meV, which indicate a highly unconventional pairing
mechanism, possibly involving multiple bands. This is very similar to the case of MgB$_2$, where two coexisting s-wave gaps were
found by the same technique. \cite{Szabo} Thermal
conductivity and specific heat measurements made by Tanatar et al. have revealed the presence of uncondensed electrons, which can
be explained by an extreme multi-band scenario, with a d-wave superconducting gap on the heavy-electron sheets of the Fermi
surface and a negligible gap on the light, three-dimensional pockets.\cite{Tanatar} 

The presence of multibands (gaps) together with the reported field dependence of the cyclotron 
masses \cite{Settai} point towards a possible temperature and/or
field dependent anisotropy in the superconducting state, which according to the standard anisotropic Ginzburg-Landau (GL) 
theory is given by
$\gamma\equiv\sqrt{m^*_{c}/m^*_{a}}=H_{c2}^{||c}/H_{c2}^{||a}=\lambda_c/\lambda_a=\xi_a/\xi_c$  ($c$ and $a$ are 
crystallographic axes, and $m, H_{c2},\lambda,$
and $\xi$ are the effective mass, upper critical field, penetration depth, and coherence length, respectively). Specifically,
in the multiband scenario proposed by Rourke et al.
\cite{Rourke} and Tanatar et al.,\cite{Tanatar} different gaps may  behave differently in magnetic field, which may lead to a
field dependent
$\gamma$. Also an anisotropic gap may  result in a temperature dependent $\gamma$.
Reports up  to date give values of the anisotropy of CeCoIn$_5$ in the range 1.5 to 2.47. For example, Petrovic et al.
have reported an anisotropy  of at least 2, as estimated from the ratio of the upper critical fields H$_{c2}$ along the $c$ 
and $a$ directions.
\cite{Petrovic}  Measurements of H$_{c2}(\theta)$ at 20 mK give a value for the anisotropy of about 2.47.\cite{Murphy}  
Magnetization measurements of the lower critical field for temperatures between 1.5 and 2.1 K give the ratio of 
the out-of-plane and
in-plane penetration depth 
$\lambda_{c}/\lambda_{a}\approx 2.3$ and of the in-plane and out-of-plane coherence length 
$\xi_{a}/\xi_{c}\approx 1.5$, which gives an anisotropy of 2.3 and 1.5, respectively.
\cite{Majumdar} 

Magnetic torque is a sensitive tool for probing the anisotropy. It has been successfully applied to investigate
the highly anisotropic high temperature superconductors and also the less anisotropic
materials such as MgB$_2$. \cite{Takahashi,Angst}  But all these previous torque measurements were made on materials which
 have negligible paramagnetism. On the other hand, CeCoIn$_5$ is a magnetic superconductor, 
so it
may have large paramagnetism which cannot be ignored in the study of the mixed state.  Here, we report torque measurements on
single crystals of CeCoIn$_5$ both in the normal state and the superconducting state. 
 Our
results show large paramagnetism in this material in the normal state. Therefore, we assume that there  are two contributions to
the torque  signal in the mixed state of  CeCoIn$_5$ single crystals: one coming from  paramagnetism and
the other one coming  from vortices. We determined $\gamma$ from the reversible part of the vortex signal and found that 
$\gamma$ is not a constant, 
instead, it is field and temperature dependent. This provides evidence that the picture in
this unconventional superconductor is not a simple single band scenario, supporting the conclusions of Rourke et al.
\cite{Rourke} and Tanatar et al.\cite{Tanatar}

\subsection{Experimental Details}
High quality single crystals of CeCoIn$_5$ were grown using a flux method. The superconducting transition temperature $T_c$,
defined as the value of $T$ at zero resistivity, is 2.3 K. The surfaces of the single crystals were etched in concentrated HCl 
for several hours and then
thoroughly rinsed in ethanol in order to remove the indium present on the surface. The mass of the single crystal for which
the data are shown is 0.75 mg. Angular dependent measurements of the magnetic  torque
experienced by the sample of magnetic moment $M$ in an applied magnetic field
$H$, were performed over a temperature range 1.9  K
$\leq T\leq$ 20  K  and applied magnetic field range 0.1 T $\leq H \leq $ 14 T using a piezoresistive torque magnetometer. 
In this 
technique, a piezoresistor measures the torsion, or twisting, of the torque lever about its symmetry axis as a result of 
the magnetic 
moment of the single crystal. The sample was rotated in the applied magnetic field between
$H
\parallel c$-axis ($\theta = 0^o$) and $H
\parallel a$-axis ($\theta = 90 ^o)$  and 
the torques $\tau_{inc}$ and $\tau_{dec}$ were measured as a function of increasing  and decreasing angle, respectively, under
 various  temperature - field conditions.

The contributions of the gravity and puck to the total torque signal were measured and subtracted from it.
 To measure the background torque due to gravity, we measured the torque signal at different
temperatures in zero applied magnetic field with the sample mounted on the puck. The gravity torque is almost temperature 
independent
and it is negligible at high applied magnetic fields.  However, as the applied magnetic
field decreases, the total torque signal becomes smaller and the effect of gravity becomes important, hence should be 
subtracted from
the measured torque. To determine the contribution of the puck to the measured torque, we measured the torque without the 
single crystal on the puck at
 different magnetic fields and temperatures. The magnitude of the torque of the puck increases with increasing magnetic field.
 Also, the contribution of the puck to
the measured torque is much larger than the contribution of the gravity. Therefore, the former contribution should always be 
subtracted from the total measured
torque signal.

\subsection{Results and Discussion}

Previous torque studies of Tl$_2$Ba$_2$CuO$_{6+\delta}$ \cite{Bergemann} and  MgB$_2$\cite{Takahashi, Angst} systems have 
shown that the normal state
torque coming from paramagnetism is small compared with the flux-vortex torque, therefore, one could neglect the former 
contribution to the total torque signal
measured in the superconducting state. However, this is not the case for CeCoIn$_5$, for which the paramagnetic torque signal 
in the normal state is comparable
with the total torque signal measured in the superconducting state, as shown below. Hence, one needs to subtract the former
 signal from the latter one in order to determine the
torque due to vortices. This is similar to the case of the electron-doped high-T$_c$ cuprate Nd$_{1-x}$Ce$_{x}$CuO$_{4}$, where a 
large paramagnetic contribution from Nd ions is discussed seperately from a superconducting contribution.\cite{Yamamoto}
Therefore, we first discuss the field and temperature dependence of the magnetic torque in the normal  state and show how we 
subtract this contribution from the measured torque in the mixed state, and then we return to the discussion of the torque signal
in the  mixed state and to the determination
of the field and temperature dependence of the bulk anisotropy.

All the torque curves measured in the normal state, some of which are shown in Fig 1, are perfectly sinusoidal and can be 
well fitted with

\begin{equation}
\tau_p(T,H,\theta) = A(T,H) \sin2\theta,
\end{equation}
where $A$ is a temperature and field dependent fitting parameter.
Indeed, note the excellent fit of the data of Fig. 1 with Eq. (1) (solid lines in the figure). The field dependence of $A/H$ 
at 1.9, 6, 10 and 20 K is shown in Fig. 2. 
The solid lines are linear fits to the data, which show that $A/H$ is linear in $H$ with a negligible $y$-intercept and a 
slope which
increases with decreasing $T$. So
$A$ is proportional to $H^2$, i.e.,
\begin{equation}
A(T,H) = C(T)H^{2},
\end{equation}
with C a temperature dependent fitting parameter.

Next we show that the torque measured in the normal state and given by Eq. (1) is a result of the paramagnetism. Indeed, the 
torque of a sample of magnetic moment
$M$ placed in a magnetic field
$H$ is given by
\begin{equation}
\vec{\tau}_p(T,H) = \vec{M}\times \vec{H}.
\end{equation}
The resultant magnetic moment $\vec{M}$ can always be decomposed into a component parallel $M_{\parallel}$ and one perpendicular 
$M_{\perp}$
to the $ab-$plane of the single crystal.  With
the magnetic field $H$  making an angle $\theta$ with the $c-$axis of the single crystal, Eq. (3) becomes:

\begin{equation}
\vec{\tau}_p(T,H,\theta) = [M_{\parallel}H\cos\theta-M_{\perp}H\sin\theta]\hat{k}.
\end{equation}
On the other hand, the experimental relationship of the torque, given by Eq. (1) with the fitting parameter $A$ given by 
Eq. (2), becomes
\begin{equation}
\tau_p(T,H,\theta) = A(T,H)\sin2\theta=2C(T)H^{2}\sin\theta\cos\theta.
\end{equation}
Therefore, with $C\equiv(C_{1}-C_{2})/2$, Eqs. (4) and (5) give
\begin{eqnarray}
M_{\parallel}=C_1H\sin\theta \equiv\chi_{a} H_{\parallel},  \nonumber  
\end{eqnarray}
where $C_{1}\equiv\chi_{a}$, the  $a$-axis susceptibility, and
\begin{eqnarray}
M_{\perp}=C_2H\cos\theta \equiv\chi_{c} H_{\perp},    
\end{eqnarray}
where $C_{2}\equiv\chi_{c}$, the $c$-axis susceptibility.
This shows that the torque measured experimentally is of the form:
\begin{equation}
\tau_p(T,H,\theta) = \frac{\chi_{a}-\chi_{c}}{2}H^2\sin2\theta
\end{equation}
The fact that $A/H = (\chi_{a}-\chi_{c})H/2\equiv (M_{\parallel}-M_{\perp})/2$ [see Eqs. (5) and (7)] shows that $A/H$ plotted in
Fig. 2 reflects the  anisotropy of the magnetic
moments  along the two crystallographic directions, $a$ and
$c$, while its linear field dependence shows that the magnetic moments are linear in $H$, hence the susceptibilities along 
these two directions are
field independent. The temperature dependence of the magnetic moments is given by the temperature dependence of the parameter 
$C$. Therefore, Eq. (7)
shows that the
$T$,
$H$, and $\theta$ dependences of the torque measured in the normal state of CeCoIn$_5$ reflect its paramagnetism and  the 
anisotropy of its
susceptibility $\Delta\chi \equiv \chi_{a}-\chi_{c}$ along the $a$ and $c$ directions.

To further check the consistency of the data  and to precisely determine the paramagnetic value of the torque, we also 
measured the magnetic moment $M$ of
the same single crystal of CeCoIn$_5$ using a superconducting quantum interference device (SQUID) magnetometer. The magnetic 
moments measured at 4, 6, 10,
15, and 20 K are plotted in the main panel of Fig. 3 as a function of the applied magnetic field for both $H\parallel c$-axis 
($\theta = 0^{o}$) and $H\parallel$ $a$-axis
($\theta = 90^{o}$). The magnetic moments for  both field orientations are linear 
in $H$ with $M_{\parallel} < M_{\perp}$ for all temperatures measured, consistent with
the torque data of Fig. 2 and with Eq. (6).
Note that the units for the magnetic moments of Figs. 2 and 3 are different. We change the units and compare the results 
of the two types of measurements. For example, the torque measured at 6 K and 5 T gives $A/H = -1.42\times 10^{-7}$ Am$^{-1}$.  
Since $A/H =(M_{\parallel}-M_{\perp})/2$, $\Delta M \equiv M_{\parallel}-M_{\perp}= -2.84\times 10^{-7}$ Am$^{-1}$. The
SQUID measurements give at the same temperature and applied magnetic field $M_\perp =  7.2\times 10^{-4}$ emu = $7.2\times 
10^{-7}$ Am$^{-1}$ and
$M_\parallel = 4.2\times 10^{-4}$ emu = $4.2\times 10^{-7}$ Am$^{-1}$; hence, an anisotropy $\Delta M =
-3.03\times 10^{-7}$ Am$^{-1}$. Therefore, the values of the magnetic moments obtained in the two types of measurements are 
within $5\%$ of each other, an
error well within  our experimental error. 

The Inset to Fig. 3 is a plot of the susceptibilities along the two directions, 
calculated from the slopes of $M(H)$ of Fig. 3. Clearly $\chi(T)$ shows anisotropy with respect 
to the field orientation. Note 
that the 
susceptibilities for both directions increase with decreasing temperature. The continuous increase of $\chi(T)$ in 
the investigated temperature range  may be related with the
non-Fermi liquid behavior due to the proximity to the quantum critical field. \cite{Stewart} These values of
$\chi_{c}$ and $\chi_{a}$ are consistent with the ones reported by other groups. \cite{Kim}  

The above study of the magnetic torque in the normal state has shown that the contribution of paramagnetism 
to the torque signal is very large, it has a quadratic $H$ dependence, and also a $T$ dependence [see Eqs. (1) and (2)]. 
Therefore, to extract the vortex torque in the
mixed state,  one needs to account for this paramagnetic contribution and subtract the two background contributions from 
the measured torque. The gravity and puck
contributions to the measured torque were determined and subtracted as explained in the Experimental Details Section. The
resultant torque  includes the paramagnetic $\tau_p$ and the vortex $\tau_v$ contributions and is plotted in the inset to Fig. 4.
We need to mention here that the vortex and  paramagnetic torque contributions have opposite signs since the magnetic moment
representing the vortex torque is diamagnetic.

The Inset to Fig. 4 is the angular dependent torque measured in the mixed state at 1.9 K and 0.3 T
in increasing and decreasing angles. Again, this torque includes 
$\tau_p$ and $\tau_v$. Note that
$\tau(\theta)$ displays hysteresis. This hysteretic  behavior is similar to the behavior in high $T_{c}$ superconductors and is a
result of intrinsic pinning.
\cite{Ishida} 
The reversible component of the torque 
is determined as the average of the torques measured in increasing and decreasing angle; i.e.,
\begin{equation}
\tau_{rev}=(\tau_{dec}+\tau_{inc})/2
\end{equation}
A plot of $\tau_{rev}(\theta)$, obtained from $\tau_{inc}(\theta)$ and $\tau_{dec}(\theta)$ data is
shown in the main panel of Fig. 4. The reversible component of the torque reflects equilibrium  states, hence it allows the 
determination of thermodynamic parameters. In
the three-dimensional anisotropic London model in the mixed state, the vortex torque $\tau_v$ is given by Kogan's 
model.\cite{Kogan} We assume that the paramagnetic contribution in the mixed state is given by Eq. (1). Therefore, 
\begin{equation}\tau_{rev}(\theta) = \tau_{p}+\tau_{v} = a\sin2\theta + \frac{\phi_0 H V}{16 \pi \mu_0 \lambda^{2}_{ab}} 
\frac{\gamma^{2}-1}{\gamma} \frac{\sin 2 \theta}
{\epsilon(\theta)} \ln\left\{\frac{\gamma \eta H^{||c}_{c2}}{H\epsilon(\theta)}\right\}
\end{equation}
where $a$ is a fitting parameter, $V$ is the volume of the sample, $\mu_{0}$
is the vacuum permeability, $\lambda_{ab}$ is the penetration depth in the $ab-$plane, $\gamma = \sqrt{m_{c} / m_{a}}$, 
$\epsilon(\theta) = (\sin^{2}
\theta+\gamma^{2}\cos^{2}\theta)^{1/2}$, 
$\eta$ is a numerical parameter of the order of unity, and
$H^{||c}_{c2}$ is the upper critical field parallel to the $c$-axis [$H^{||c}_{c2} (1.9$ K$) = 2.35$ T]. 
We define $\beta \equiv \phi_0 HV/(16\pi  \mu_0 \lambda^{2}_{ab})$. 

To obtain the field dependence of the anisotropy $\gamma$, we fit the torque  data with Eq. (9), with $a$, $\beta$, and $\gamma$
as three fitting parameters. The solid line in the main panel of Fig. 4 is the fitting result for $T$ = 1.9 K
and $H$ = 0.3 T. We need to mention here that the value of the fitting paramter $a$ is 20$\%$ smaller than what we would expect 
from the extrapolation of the normal state paramangetic torque data. Also, $a$ has an $H$ dependence with an exponent of 2.3
instead of 2.  So, either there is extra contribution from other physics which has a weak field dependence, in addition to the
paramagnetism contribution, or maybe the paramagnetic contribution becomes smaller in the mixed state of CeCoIn$_5$. Further
experiments are needed to clarify this issue.  

Figure 5 is a plot of the field dependence of $\beta$. The Inset
is an enlarged plot of the low field region. Note that $\beta$ displays linear behavior up to a certain field with no
y-intercept, then it deviates from  linearity at $H
\approx 0.5$ T, and it increases fast in the high field region. Since, on one hand $\lambda$ should be field
independent, and on the other hand Eq. (9) is valid only for applied magnetic fields much smaller than the upper critical field,
i.e.
$H << H_{c2}(T)$ for a given temperature, we assume that  $0.5$ T, the field
at which
$\beta(H)$ deviates from linearity, is the cutoff field $H_{cut}$ for the applicability of the above theory. The slope of
$\beta(H)$ in the linear $H$ regime gives $\lambda$ ($T$ = 1.9 K) = 787 nm. This value is larger than previous reports, which give
$\lambda_{ab}$ = 600 nm from measurements using a tunnel diode oscillator \cite{Ozcan} and $\lambda_{ab}$ = 330 nm from
magnetization measurements.
\cite{Majumdar}

Next, we fix $\lambda$ to the three values given above and fit the $\tau_{rev}(\theta)$ data with
only two fitting parameters,
$a$ and $\gamma$. The resultant  field dependence of
$\gamma$ is shown in Fig. 6 for the different $\lambda$ values. The parameter $\gamma$ is first decreasing with increasing $H$, 
reaches a minimum at $H$ = 0.5 T, and then increases with further increasing field. As mentioned above, the cutoff field is 0.5
T. The data for $H>H_{cut}$ are not reliable due to the failure of Kogan's theory in this $H$
region. So we conclude that the anisotropy $\gamma$ decreases with increasing field. We note that this field dependence of
$\gamma$ in CeCoIn$_5$ is opposite with the one for MgB$_2$, \cite{Angst} in which $\gamma$ increases with increasing field. We
found that the value of $\gamma$ is very sensitive to the value of $\lambda$, i.e., the larger the value of
$\lambda$, the larger the value of $\gamma$, with no effect however on its $H$ dependence. 

To study the temperature dependence of $\gamma$, we performed torque measurements in the mixed state at 1.9 K, 1.95, 2.00 K in an 
applied magnetic field of 0.3
T, and determined $\gamma$ from Eq. (9). Figure 7 is a composite plot of the temperature dependence of the anisotropy. The 
squares give $\gamma(T)$ determined from
torque measurements as discussed above [with $\lambda $ (1.9 K) = 600 nm, $\lambda $ (1.95 K) = 670 nm, and $\lambda $ (2 K) = 740
nm taken from Ref (25)], the solid circles give $\gamma(T)$ calculated from the ratio of
$H_{c2}^{||c}/H_{c2}^{||a}$ taken from previous reports,\cite{Tayama} while the triangles give $\gamma(T)$ determined from 
$\sqrt{\rho_{c}(T)/\rho_{a}(T)}$ measured
in zero field [see $\rho_c(T)$ and $\rho_{a}(T)$ in Inset to Fig. 7]. Note that the overall trend is a decrease of the 
anisotropy with increasing $T$, with a
stronger dependence around $T_{c}(0.3$ T$)=2.23$ K. The 
values of $\gamma$ obtained from torque measurements are well within the range previously reported. The fact that the anisotropy
depends both on temperature and magnetic field could explain its relatively wide range of values reported in the literature.

 The field and temperature dependence of the anisotropy implies the breakdown of the standard anisotropic GL theory, 
which assumes a single band anisotropic
system with a temperature and field independent effective-mass anisotropy. In fact, a temperature and field
dependent anisotropy was previously observed in NbSe$_2$,\cite{Muto} LuNi$_2$B$_2$C,\cite{Metlushko} and MgB$_2$,\cite{Ferdeghini, 
 Angst} the latter one having a value of $\gamma$ similar with
CeCoIn$_5$. Very recently, Fletcher et al. reported measurements of the temperature-dependent anisotropies 
($\gamma_{\lambda}$ and $\gamma_{\xi}$) of both the London penetration depth $\lambda$ and the upper critical field of
MgB$_2$.\cite{Fletcher} Their main result is that the anisotropies of the penetration depth and $H_{c2}$ in MgB$_2$ have opposite
temperature dependences, but close to $T_{c}$ they tend to a common value [see Fig. 4 of Ref (30)]. This result confirms nicely
the theoretical calculation of anisotropy based on a two-gap scenario.\cite{VGKogan} So, the temperature dependence of $\gamma$ in
CeCoIn$_5$ and MgB$_2$ could have a similar origin. The field dependence of $\gamma$ could also be related with a multiband
structure. In this scenario,  the response to an applied magnetic field may be different in the case of the d-wave
superconducting gap on the heavy-electron sheets of the Fermi surface and the negligible gap on the light, three-dimensional
pockets.

\subsection{Summary}
Torque measurements were performed on CeCoIn$_5$  single crystals in both the superconducting and normal states. Two 
contributions to the torque signal in the
mixed state were identified: one from paramagnetism and the other one from the vortices. The torque curves show sharp hysteresis
peaks at
$\theta = 90^{o}$  ($\theta$ is the angle between
$H$ and the
$c-$axis of the crystal) when the measurements are done in clockwise and counterclockwise directions. This hysteresis is a 
result of 
the intrinsic pinning of vortices, a behavior very similar to high transition temperature cuprate superconductors. The 
temperature and magnetic field dependence of
the anisotropy $\gamma$ is obtained from the reversible part of the vortex torque. We find that $\gamma$ decreases
 with increasing magnetic field and
temperature. This result indicates the breakdown of the Ginzburg-Landau theory, which is based on a single band model and 
provides evidence for a multiband picture
for CeCoIn$_5$, which is highly possible due to its complex Fermi topology. \cite{Hall, Settai}
\\
\\
\textbf{Acknowledgments}

The authors would like to thank Vladimir Kogan for fruitful discussions. This research was supported by the National Science 
Foundation under Grant
No. DMR-0406471 at KSU and the US Department of Energy under Grant No.
DE-FG02-04ER46105 at UCSD. \label{} 
\\


\section{Figure Captions}
Figure 1. (Color online). Angular $\theta$ dependence of the paramagnetic torque $\tau_p$ measured in the normal state of
CeCoIn$_5$ at  different temperatures $T$ and  applied magnetic field $H$ values. The 
solid lines are fits of the data with Eq. (1). Inset: Schetch of the single crystal with the orientation of the magnetic field
$H$ and torque $\tau$ with respect to the crystallographic axes.

Figure 2. (Color online). Field $H$ dependence of $A/H$, where $A$ is the fitting parameter in Eq. (1). The solid lines are linear
fits  of the data. 

Figure 3. Plot of the magnetic moment $M$ vs applied magnetic field $H$, with $H\parallel c$-axis (solid symbols) and 
$H \parallel a-$ axis (open symbols), measured
at 4, 6, 10, 15, and 20 K. Inset: Susceptibility $\chi$ vs temperature $T$, measured with $H\parallel c$-axis and 
$H \parallel a$-axis. 

Figure 4. (Color online). Angular $\theta$ dependence of the reversible torque
$\tau_{rev}$, measured in the mixed state of CeCoIn$_5$ at a temperature $T$ of 1.9 K and an applied magnetic field 
$H$ of 0.3 T. The solid line is a fit of
the data with Eq. (9). Inset: $\theta$ dependence of the hysteretic torque $\tau$, measured in increasing 
and decreasing angle at the same $T$ and $H$. 

Figure 5. Magnetic field $H$ dependence of the fitting parameter $\beta$. The solid line is a guide to
the eye. Inset: Enlarged plot of the low field region of the data in the main panel.

Figure 6. (Color online). Field $H$ dependence of the anisotropy $\gamma$ measured at 1.9 K.

Figure 7. (Color online). Composite plot of the temperature $T$ dependence of the anisotropy $\gamma$. The circles
show the results obtained from the upper critical field data [Ref. 26], the triangles are obtained from the resistivity
data shown in the Inset of this figure, and  the squares  are from the present torque data measured in an applied
magnetic field of 0.3 T. Inset: $T$ dependence of the in-plane $\rho_{a}$ and out-of-plane $\rho_c$ resistivities measured 
in zero field.

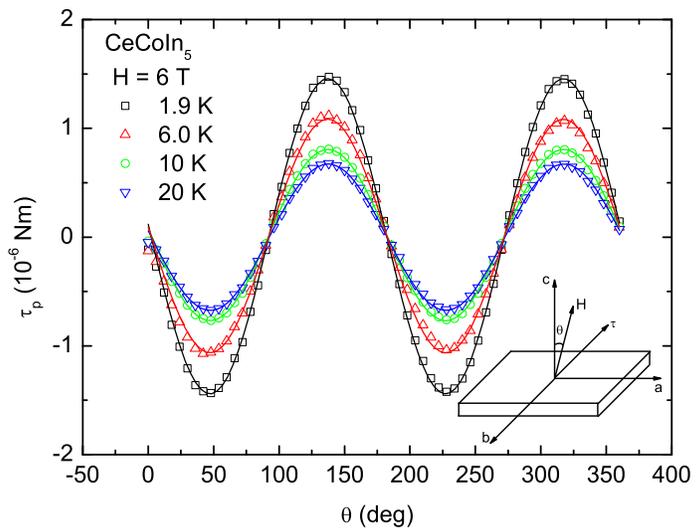
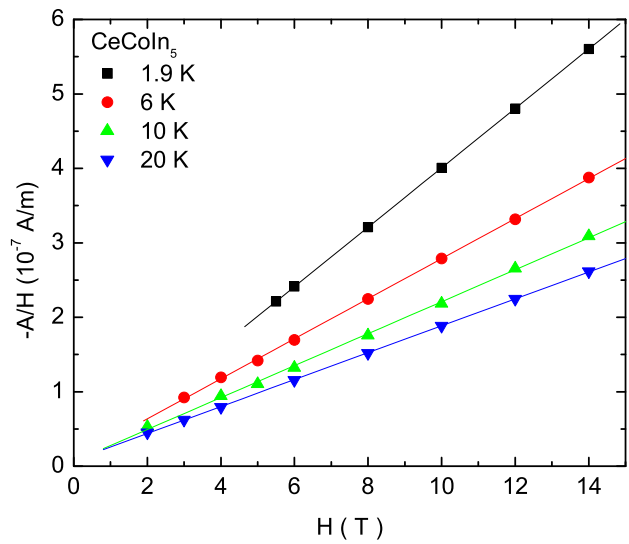
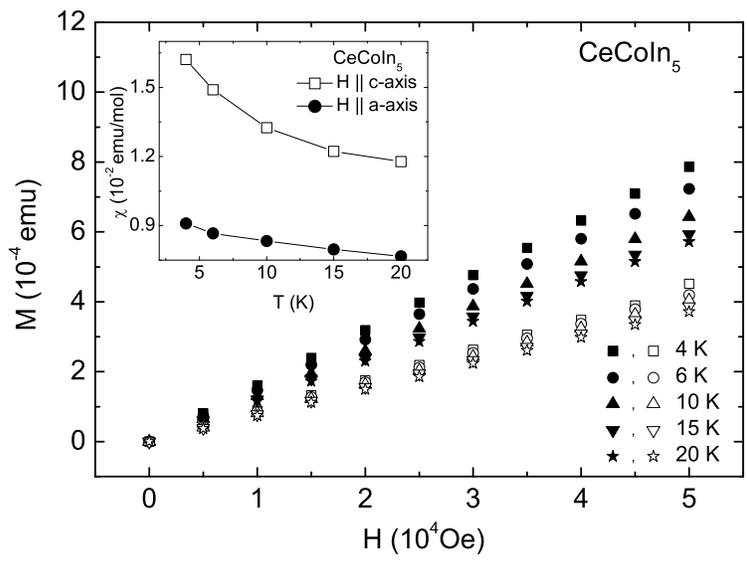
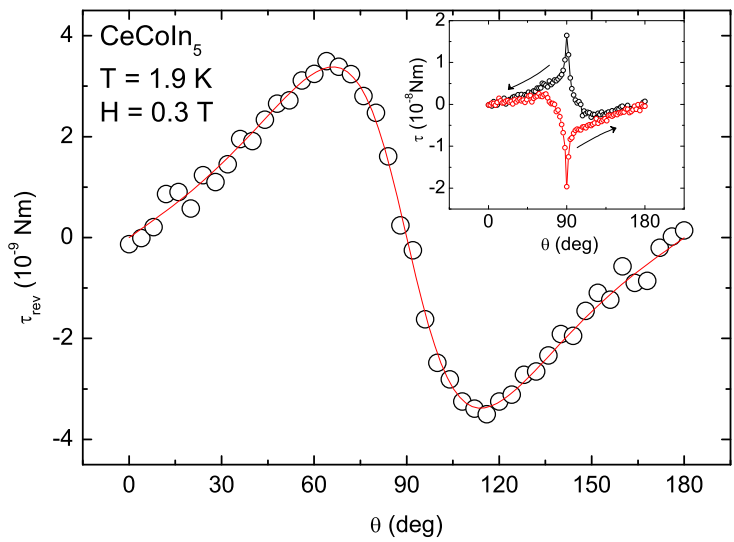
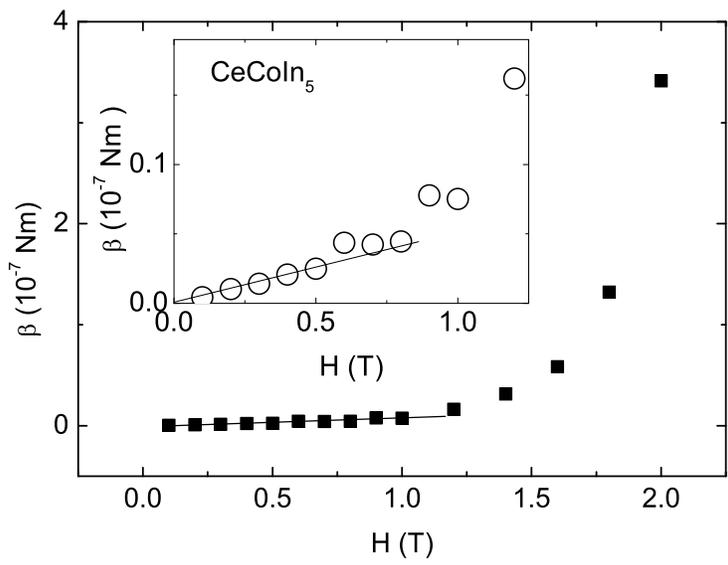
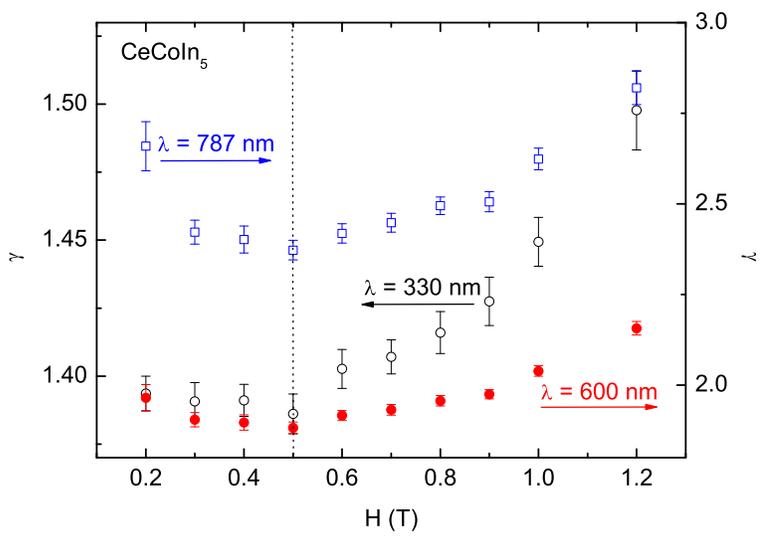
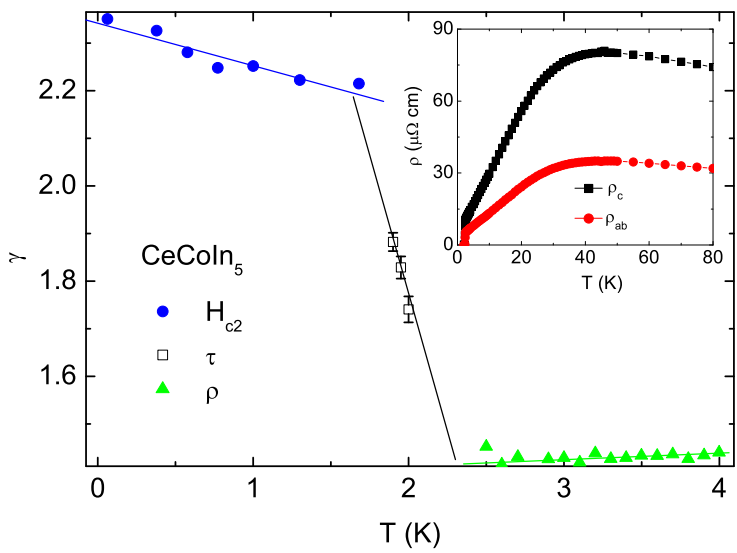
\end{document}